\documentclass[aip,jcp,preprint]{revtex4-1}
\usepackage{xcolor}
\usepackage{microtype}
\usepackage{hyperref}
\usepackage{graphicx}

\begin{document}

\title{Poisson ratio and excess low-frequency vibrational states in glasses}

\author{Eug\`{e}ne Duval}
\affiliation{Universit\'e de Lyon, F-69622, France; Universit\'e Lyon 1,
Institut Lumi\`ere Mati\`ere, CNRS, UMR 5586,
F-69622 Villeurbanne Cedex, France}

\author{Thierry Deschamps}
\affiliation{Universit\'e de Lyon, F-69622, France; Universit\'e Lyon 1,
Institut Lumi\`ere Mati\`ere, CNRS, UMR 5586,
F-69622 Villeurbanne Cedex, France}

\author{Lucien Saviot}
\affiliation{Laboratoire Interdisciplinaire Carnot de Bourgogne,
UMR 6303 CNRS-Universit\'e de Bourgogne,
9 Av.
A.
Savary, BP 47 870, F-21078 Dijon Cedex, France}
\date{\today}

\begin{abstract}
In glass, starting from a dependence of the Angell's fragility on the Poisson ratio
[V. N. Novikov and A. P. Sokolov, Nature 431, 961 (2004)],
and a dependence of the Poisson ratio on the atomic packing density
[G. N. Greaves \textit{et al.}, Nat. Mater. 10, 823 (2011)],
we propose that the heterogeneities are predominantly density fluctuations in \emph{strong} glasses (lower Poisson ratio) and shear elasticity fluctuations in \emph{fragile} glasses (higher Poisson ratio).
Because the excess of low-frequency vibration modes in comparison with the Debye regime (boson peak) is strongly connected to these fluctuations, we propose that they are breathing-like (with change of volume) in \emph{strong} glasses and shear-like (without change of volume) in \emph{fragile} glasses.
As a verification, it is confirmed that the excess modes in the \emph{strong} silica glass are predominantly breathing-like.
Moreover, it is shown that the excess breathing-like modes in a \emph{strong} polymeric glass are replaced by shear-like modes under hydrostatic pressure as the glass becomes more compact.
\end{abstract}

\maketitle

\section{Introduction: Poisson ratio and fragility}

The Poisson ratio ($\nu$) is the negative quotient of the transverse strain and the axial strain of a solid under a uniaxial stress.
It is related to the bulk modulus $K$ and the shear modulus $G$ as follows:

\begin{equation}
\label{eq.1}
\nu=\frac{3K-2G}{2\left(3K+G\right)}
\end{equation}
From the relation that exists between the elastic moduli and the velocities of the longitudinal ($V_\ell$) and transverse ($V_t$) acoustic waves, one obtains:

\begin{equation}
\label{eq.2}
\nu=\frac{\left(\frac{V_\ell}{V_t}\right)^2-2}{2\left[\left(\frac{V_\ell}{V_t}\right)^2-1\right]}
\end{equation}

Because both $K$ and $G$ have positive values, $\nu$ can take values in between $\frac12$ and -1.
The values which are close to $\frac12$ are obtained for $K \gg G$ which means that the material is not very compressible and is more easily strained by a shear stress.
It is the contrary for $\nu$ close to 0.
In the case of negative values ($-1<\nu<0$) the material swells under a tension.
Such a strange behavior is observed for so-called auxetic materials.
The mechanical properties of non-crystalline materials, and especially of glasses, were well described in relation to the Poisson ratio by Greaves \textit{et al.} in a recent paper.\cite{greaves,*greavesCOR}
After the early paper of Makishima and Mackenzie,\citep{mack}
these authors established a relation between the Poisson ratio and the atomic packing density which is defined as the ratio of the minimum theoretical volume occupied by the ions to the corresponding effective volume in the glass.
A packing ratio close to one corresponds to $\nu \approx{0.5}$ and a low packing ratio corresponds to $\nu \approx 0$.

Glass-forming materials are often characterized by the Angell's fragility which measures the deviation from the Arrhenius regime of the temperature dependence of the shear viscosity ($\eta$) at the glass transition temperature ($T_g$).
Angell's fragility is quantified by the index $m$ as defined in Eq.~\ref{angell}.

\begin{equation}
\label{angell}
m = \left[\frac{\partial~\log~\eta}{\partial~T_{g}/T}\right]_{T=T_g}
\end{equation}

Novikov \textit{et al.} evidenced the relation which exists between the Angell's fragility of the glass-former and the Poisson ratio of the corresponding glass.\cite{novikov2004,novikov2005}
They showed that the fragility $m$ is linearly dependent on the ratio $K/G$.
Yannopoulos and Johari \cite{yann}, considering a large number of glasses, found that this linear dependence is not general and is only valid for a limited number of glasses.
In fact, more generally, $m$ increases with $K/G$ and the slope of the approximate linear dependence changes from one type of glass to another.\cite{greaves}
From Eq.~\ref{eq.1}, it means that, considering a type of glass, the Poisson ratio is weaker for glasses froze-in from strong melts than for glasses frozen-in from fragile melts.
Obviously, the Poisson ratio of the glass is correlated to the Poisson ratio of the melt.
Indeed, it was experimentally demonstrated \cite{rouxel} that the Poisson ratio of the considered fragile glass-formers strongly decreases when the temperature approaches the glass-transition one (T$_{g}$), while that of strong glass-formers does not change very much when the temperature is approaching T$_{g}$.
It means that the difference between the Poisson ratios of fragile and strong glass-formers is even larger than that between the Poisson ratios of the corresponding fragile and strong glasses.
This interesting experimental result justifies that, in the following, glasses with a low Poisson ratio (typically, $\nu<0.2$) will be called \emph{strong} and those with a high Poisson ratio (typically, $\nu>0.25$) will be called \emph{fragile}.
From  the relation between the Poisson ratio and the atomic packing density,\cite{greaves} it means that the fragile glasses are more compact than strong ones.

 
\section{Nature of the vibrational modes in the boson peak related to the Poisson ratio}

It is now generally accepted that the elasticity of glasses at the nanometric scale is heterogeneous.
This was early hypothesized in a simple model relating the frequency of boson peak to the mean size of the heterogeneities.\cite{duval90}
Later theoretical and computational models confirmed that shear inhomogeneities can account for the boson peak.
See the recent paper of Marruzzo \textit{et al.}\cite{marruzzo} and the numerous references therein for more about these inhomogeneities.
The reason for the heterogeneous elasticity (not only for shear) is either the fluctuations of atomic packing density or the fluctuations of the shear elasticity which is not necessarily correlated to density fluctuations.
One understands that vibrations can at least partially be localized around the such heterogeneities, whatever their origin.
This localization is responsible for the low-frequency vibrational density of states (VDOS) excess, in comparison with the Debye regime, that is observed as the so-called boson peak.
The VDOS excess does not consist of only the localized modes themselves.
It is due to the hybridization between the partially localized modes and the propagating longitudinal or transverse acoustic modes.\cite{duval2007}
At the anti-crossing between the dispersive acoustic branches and the non-dispersive pseudo-optical branches corresponding to the localized modes there is a flattening of the dispersion curves, that gives  rise to pile-ups in the VDOS at the origin of the boson peak.

Two different types of localized vibration modes are distinguished: the modes inducing a volume change, and the modes without volume change.
The first ones are breathing-like modes or bulk-like modes, and the second ones are shear modes.
The frequency of the bulk-like mode ($\omega_b$) is well-approximated by the following equation: $\omega_b=\frac{V_b}{D}$, $V_b$ being the bulk velocity and $D$ the length scale of  heterogeneity.
The frequency of the localized shear mode is $\omega_t \approx \frac{V_t}{D}$.
The bulk velocity is proportional to the square root of the bulk modulus so that:

\begin{equation}
\label{eq.3}
V_b^2=V_\ell^2-\frac43 V_t^2
\end{equation}

From these equations, we note that the ratio $\frac{\omega_b}{\omega_t}$ increases with increasing $\frac{V_\ell}{V_t}$ and therefore with increasing $\nu$.
This is consistent with the fact that the glass becomes more compact when $\nu$ increases.
From the above dependencies of the vibration frequencies on velocities, $\omega_b=\omega_t$ for $\nu=0.13$.
A more rigorous calculation for modes localized on a perfectly spherical heterogeneity\cite{lamb} indicates that $\omega_b=\omega_t$ closer to $\nu=0$.
In any case, it follows that for glasses frozen-in from strong melts ($\nu \approx 0.13$) $\omega_b$ is close to $\omega_t$.
Of course, having the lowest frequency is not a sufficient criterion to determine the contribution of the modes to the excess in the VDOS.
A more decisive parameter is the strength of the localization and the resulting hybridization between  the localized vibrational modes and the acoustic propagating modes.
The modes are all the more localized when the fluctuations, which cause the localization, are marked.
It was noted before that a weak Poisson ratio ($\nu \approx 0.13$) is related to the prevalence of atomic volume (or density) fluctuations .
On the other hand, a high Poisson ratio ($\nu \approx 0.35$) is linked to the existence of shear fluctuations without change of atomic volume.
From these considerations, we expect that in a strong glass  with a  weak $\nu$ the VDOS excess is dominated by the breathing-like vibrations (with change of volume).
On the other hand, in a glass with a higher $\nu$, the VDOS excess is dominated by shear vibration modes (without change of volume).

\subsection{Excess vibrational states in a polymeric glass.
Effect of hydrostatic pressure}

The experimental results on the shift of the boson peak frequency under applied hydrostatic pressure  obtained by Stavrou \textit{et al.} with a polymeric glass are very interesting.
They were very recently published.\cite{stavrou,*stavrouCOR}
The polymeric glass considered in this work, Kel-F 800, is a co-polymer of chlorotrifluoroethylene and vinylidene fluoride in a 3:1 weight ratio.
Its glass transition temperature $T_g$ at ambient pressure is 26$^{\circ}$C.
By comparison with the acoustic velocities also measured by Brillouin scattering as a function of applied hydrostatic pressure by Stevens \textit{et al.},\cite{stevens} it appears (Eq.~\ref{eq.2}) that $\nu=0.166$ at ambient pressure and abruptly increases to 0.37 at 0.3~GPa.
This value then remains constant up to pressures higher than 15~GPa.
This means that the polymeric strong glass at ambient pressure, becomes fragile under an applied pressure higher than 0.3~GPa.
This change was interpreted by a collapse of free volumes,\cite{stevens} the glass becoming more compact.
Stevens \textit{et al.} noted that the co-polymer is in a glassy state at any pressure including ambient.\cite{stevens}
Regarding the glassy state, it is interesting to notice that at $P=0.1$~GPa the Poisson ratio ($\nu=0.06$) is weaker than at ambient pressure ($\nu=0.166$).
This decrease is due to the increase of $V_t$ from 1225 to 1321~m/sec as the pressure is applied without a significant change of $V_\ell$.\cite{stevens}
It is very likely that the increase of $V_t$ comes from the increase of $T_g$ and that the polymeric glass at ambient pressure is stronger than indicated by its Poisson ratio.

\begin{figure}[!ht]
 \includegraphics[width=\columnwidth]{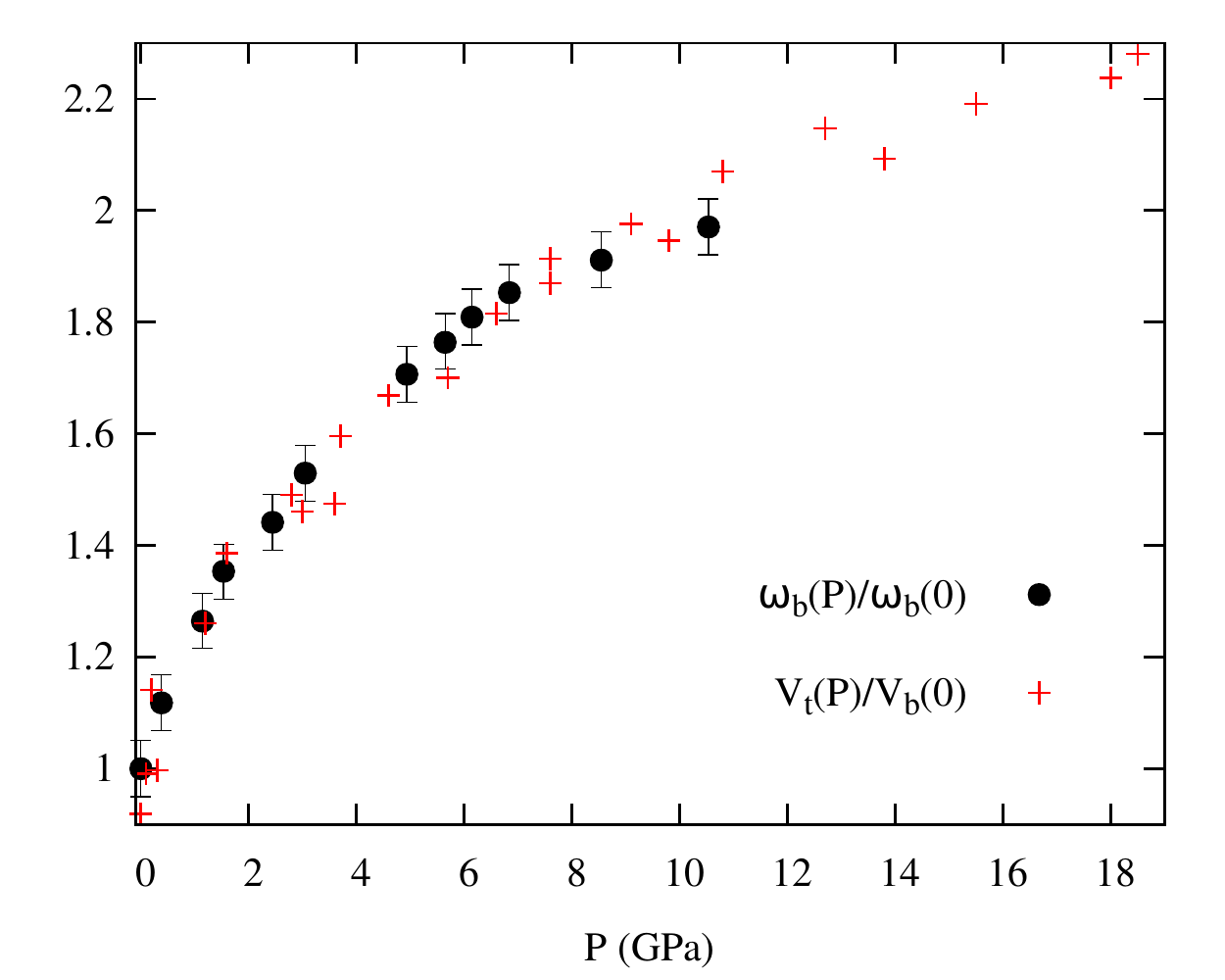}
 \caption{\label{fig} Variations with pressure of $\frac{\omega_{bp}(P)}{\omega_{bp}(0)}$ (full circles with error bars, from Stravrou \textit{et al.}\cite{stavrou}) and $\frac{V_t(P)}{V_b(0)}$ (crosses, from Stevens \textit{et al.}\cite{stevens}  (red online)) for the Kel-F 800 glass.}
\end{figure}

Stavrou \textit{et al.}\cite{stavrou,*stavrouCOR} compared the ratio of the frequency of the boson peak at pressure $P$ to the frequency of the boson peak at ambient pressure $\frac{\omega_{bp}(P)}{\omega_{bp}(0)}$ to the ratio of the bulk velocities $\frac{V_b(P)}{V_b(0)}$.
By doing so, they assumed that the frequency of the boson peak is given by $\frac{V_b}{D}$, \textit{i.e.}, that the excess modes are the bulk-like modes.
From this comparison they deduced that the length scale $D$ increases with pressure because
$\frac{\omega_{bp}(P)}{\omega_{bp}(0)}<\frac{V_b(P)}{V_b(0)}$.
Such an increase of $D$ with pressure is very surprising.
But their experimental results can be interpreted differently.
Following the previous considerations, at ambient pressure the glass is strong ($\nu=0.166$) and the bulk-like modes (with change of volume) are in excess.
On the other hand, at a pressure higher than 0.3~GPa the glass is fragile ($\nu=0.37$) and the shear modes (without change of volume) are in excess.
As a result, $\frac{\omega_{bp}(P)}{\omega_{bp}(0)}$ should be compared to $\frac{V_t(P)}{V_b(0)}$ for $P>0.3$~GPa.
This is done in Fig.~\ref{fig} in which we observe a very good agreement for $P>0.3$~GPa.
Furthermore, it was remarked that $\frac{V_t(P)}{V_t(0)}$ was systematically higher than
$\frac{\omega_{bp}(P)}{\omega_{bp}(0)}$.
This comparison shows that the modes of the boson peak, which are likely bulk-like at ambient pressure, are replaced by shear modes at a pressure higher than 0.3~GPa.
Furthermore, it comes from this comparison that the length scale $D\approx 1.3$~nm of the localized modes hardly changes with pressure.
This means that the shear fluctuations under hydrostatic pressure keep the memory of the density fluctuations at ambient pressure.

\subsection{Excess vibrational states in silica glass}

Silica is a strong glass at ambient pressure.
According to Zha \textit{et al.},\cite{zha} $\nu=0.15$.
On the other hand, the depolarization ratio of the Raman boson peak is 0.3.
If the vibration modes involved in the boson peak were the shear modes then the depolarization ratio would be closer to 0.75.\cite{novikov95}
The Poisson ratio and the measured depolarization ratio let us assign the modes of the boson peak to breathing-like modes.
Recently Ruffl\'e \textit{et al.} showed that the temperature dependence of $\omega_{bp}$ for the silica glass scales as that of $V_b$.\cite{ruff}
This is another strong indication that the modes of the boson peak are mainly breathing-like.

There are no accurate experimental measurements of the shift of the  boson peak frequency of silica with an applied  hydrostatic pressure.
However, such a shift was measured by Raman scattering \cite{thierry} for the GeO$_{2}$ glass which is also strong and has a similar behavior as SiO$_{2}$ under an applied pressure.
For GeO$_{2}$ it is noted that, while the transverse acoustic velocity presents a minimum at about 1.5~GPa, as with SiO$_{2}$ at about 3.5~GPa\cite{suito}, the shift of the boson peak frequency does not show a minimum and monotonically increases between 0 and 4 ~GPa as the bulk velocity.
The SiO$_{2}$ glass becomes fragile ($\nu=0.33$) at about 20~GPa.\cite{zha}
To the best of our knowledge, the boson peak of silica glass has not been measured at pressures higher than 20~GPa.
Our findings make it possible to theoretically extrapolate the shear character of the excess VDOS above 20~GPa.

\section{Discussion}

From this study, it is deduced that strong glasses which have a weak Poisson ratio ($\nu<0.2$) and are not compact, are affected mainly by fluctuations of atomic volume or of density.
As a result, localized breathing-like vibration modes can exist.
It is then likely that the excess vibrational states have a breathing character.
On the other hand, fragile glasses, which have a higher Poisson ratio ($\nu>0.25$) and are more compact, are affected by fluctuations of shear elasticity without change of volume.
Therefore the excess vibrational states have a transverse character because only such modes can be localized in such a system.

Applying an hydrostatic pressure to a strong glass increases its compactness and therefore its Poisson ratio.\cite{rouxel2}
As a result, a glass which is strong at ambient pressure may become fragile at higher pressure.
As a consequence, the modes in the boson peak which are breathing-like at ambient pressure are replaced by shear modes under pressure.
However, there are exceptions to this rule.
As an example, the Poisson ratio of the poly(methyl methacrylate) glass (PMMA) is changed from $\nu=0.325$ at ambient pressure to $\nu=0.125$ at P=0.275~GPa.\cite{weishaupt}
This decrease of $\nu$ is likely due to  a modification of the macromolecules arrangement becoming locally more ordered under pressure.
It would be interesting to check if the shear modes at ambient pressure are replaced by breathing-like modes in the spectral range of the boson peak when a hydrostatic pressure is applied.

It is interesting to note that there is a correlation between the nature of the excess modes and that of the relaxation at the glass transition.
Buchenau \textit{et al.}\cite{buchenau} showed that in strong glass formers (low Poisson ratio), the relaxations couple more strongly to density fluctuations.
By contrast, they tend to couple to shear fluctuations in fragile glass-formers (high Poisson ratio).

It is generally considered that the modes of the boson peak are transverse as claimed by Shintani \textit{et al.}\cite{shintani}
The examples that we have given show the opposite for strong glasses with a low Poisson ratio.
In that case, the modes in the spectral range of the boson peak can be predominantly breathing-like.
Another argument in favour of the universal transverse nature of the excess vibrational states comes from simulations.\cite{shintani}
Unfortunately, simulations are in general not performed for very strong glasses with a low Poisson ratio as in this recent paper by Marruzo \textit{et al.}\cite{marruzzo}
By comparing a mean-field theory of shear-elastic heterogeneity \cite{schirma} with a large-scale simulation of a soft-sphere glass, the authors concluded that the origin of the boson peak is the heterogeneous shear elasticity.
The system under consideration was very fragile as shown by the calculated Poisson ratio which is high ($\nu \approx{0.35}$).
This result agrees with our deduction that the heterogeneities of elasticity and the modes of the boson peak have a shear character in fragile glasses.
However, it isn't at odds with the modes of the boson peak in strong glasses being breathing-like, especially in silica glass.

It is remarkable that the VDOS excess clearly increases with the
decrease of the Poisson ratio as shown by Novikov \textit{et al.}\cite{novikov2005}
From the deductions of our study, this means that VDOS excess is more related to the
amplitude of the fluctuations of density than to that of shear
elasticity and mainly consists of breathing-like vibrational states in
strong glasses.

Monaco \textit{et al.}\cite{monaco} showed that the characteristics of the boson peak (amplitude and frequency) of a silicate glass (Na$_2$FeSi$_3$O$_{8.5}$) evolve as a function of densification according to the Debye regime.
Their conclusion was that the glass behaves like a simple elastic medium without change of the local structure with densification in the range from 2.71 to 2.88~g/cm$^3$.
We note that the glass under consideration was not strong and had a Poisson ratio $\nu \simeq 0.25$ whatever the considered densities.
We then deduce that on the one hand the excess modes are shear modes whatever the densification, and on the other hand the volume contraction after densification by the application of an hydrostatic pressure is homogeneous (no change of $\nu$).
Therefore the obtained experimental result concerning the boson peak is not surprising.
However it does not imply that the shear elasticity is that of a continuous elastic medium, and that there are not shear elasticity fluctuations.
The mere presence of a VDOS excess is proof of the contrary.

For the same silicate glass, densifying up to 3.25~g/cm$^3$ does not noticeably change the Poisson ratio and it was shown that the boson peak is equivalent to a van Hove TA singularity  which, after extrapolating to higher densities, reaches that of the crystal of the same composition.\cite{chumakov}
This result is in agreement with the hybridization model of the transverse propagating acoustic states with the shear localized ones.\cite{duval2007}
If for higher densities the boson peak eventually turns into the van Hove singularity of the crystal, it could be concluded that the shear elasticity of the glass prefigures that of the crystal of the same composition.
Recently, Baldi \textit{et al.}\cite{baldi} compared
the inelastic X-ray scattering (IXS) of ``polycrystalline'' $\alpha$-quartz (2.649~g/cm$^3$) to that of silica glass densified to the same density (2.67~g/cm$^3$).
They observed that the vibrations  of the densified glass resemble to those of the polycrystal.
However it is pointed out that for a similar densified silica glass (density=2.62~g/cm$^3$), the dispersion curve shows a cross-over\cite{foret} (flattening) at an energy $E\approx 9$~meV and a momentum-transfer $Q\approx 2.2$~nm$^{-1}$.
Such a wave-vector is much smaller than that of a van Hove singularity of $\alpha$-quartz.
It corresponds\cite{duval2007} to an heterogeneity size $D\approx 1.8$~nm which is not far from that estimated by Baldi \textit{et al.}\cite{baldi} from the width of the IXS peak.

\section{Conclusion}

This study evidences that the relevant parameter to take into account when studying the low-frequency vibrational dynamics of glasses is the Poisson ratio.
\textit{Strong} glasses have a low Poisson ratio and are not compact.
The excess low-frequency vibrational states in such glasses are predominantly breathing-like, \textit{i.e.}, with volume change.
On the other hand, \textit{fragile} glasses are more compact and not very compressible.
The excess low-frequency vibrational states are shear-like in these glasses.
Under an applied hydrostatic pressure, a \textit{strong} glass may become \textit{fragile} and the excess vibrational states evolve from breathing-like to shear-like.

\begin{acknowledgments}

The authors are very grateful to B. Champagnon and A. Mermet for discussions and suggestions, and to D. B. Murray for his critical reading of the manuscript.

\end{acknowledgments}

%

\end{document}